\documentclass[12pt]{article}
\usepackage{amsmath}
\usepackage{graphics,graphicx,epsfig}
\usepackage{amssymb}
\usepackage{epstopdf}
\usepackage{sidecap}
\usepackage{appendix}
\usepackage{xcolor}
\usepackage{subfigure}

\usepackage{caption}

\def\calt{{\cal T}}
\def\calm{{\cal M}}

\def\bhat{{\hat B}}
\def\that{{\hat t}}
\def\xhat{{\hat x}}

\def\phimax{\Phi_{max}}
\def\phitot{\Phi_{tot}}

\begin{document}

\begin{titlepage}
\vfill
\begin{flushright}
ACFI-T20-13\\
\end{flushright}

\vskip 1.2in

\begin{center}
\baselineskip=16pt
{\Large\bf Geometry of AdS-Melvin Spacetimes}
\vskip 0.15in

{\bf  David Kastor and Jennie Traschen} 

\vskip 0.4cm
Amherst Center for Fundamental Interactions\\
Department of Physics\\ University of Massachusetts, Amherst, MA 01003\\

\vskip 0.1 in Email: \texttt{kastor@umass.edu, traschen@umass.edu}
\vspace{6pt}
\end{center}
\par
\begin{center}
{\bf Abstract}
 \end{center}
\begin{quote}
We study asymptotically AdS generalizations of  Melvin spacetimes, describing gravitationally bound tubes of magnetic flux.  We find that narrow fluxtubes, carrying strong magnetic fields but little total flux, are approximately unchanged from the $\Lambda=0$ case at scales smaller than the AdS scale.  However, fluxtubes with weak fields, which for $\Lambda=0$ can grow arbitrarily large in radius and carry unbounded magnetic flux, are limited in radius by the AdS scale and like the narrow fluxtubes carry only small total flux.  As a consequence, there is a maximum magnetic flux $\phimax = 2\pi/\sqrt{-\Lambda}$ that can be carried by static fluxtubes in AdS.   For flux $\phitot<\phimax$ there are two branches of solutions, with one branch always narrower in radius than the other.  We compute the ADM mass and tensions for AdS-Melvin fluxtube, finding that the wider radius branch of solutions always has lower mass.  In the limit of vanishing flux, this branch reduces to the AdS soliton.
\vfill
\vskip 2.mm
\end{quote}
\hfill
\end{titlepage}


\section{Introduction}

The Melvin solution \cite{Melvin:1963qx} to $4D$ Einstein-Maxwell theory describes a stable \cite{Melvin:1965zza,thorne_stability} gravitationally bound tube of magnetic flux and has enjoyed enduring interest over the years in a number of physical contexts.  The solution generating tool developed in \cite{Ernst:1976mzr} allows a Melvin-type magnetic field to be added to any axisymmetric, asymptotically flat solution to the Einstein-Maxwell equations.  Recent studies of rotating black hole spacetimes immersed in magnetic fluxtubes  include \cite{Gibbons:2013yq,Gibbons:2013dna,Cvetic:2013roa}.   

Adding a magnetic field to the magnetically charged C-metric, which describes a pair of Reissner-Nordstrom black holes accelerating away from one another \cite{Kinnersley:1970zw}, produces the Ernst spacetime \cite{ernst-cmetric} where the magnetic field supplies the accelerating forces on the black holes, rather than the unphysical conical singularities of the C-metric.  The Euclidean Ernst spacetime, in turn, provides an instantonic pathway for a magnetic field to decay via pair production of magnetically charged black holes \cite{Garfinkle:1993xk}.  

Generalizations of the Ernst spacetime in dilaton gravity theories were found in \cite{Dowker:1993bt}.  It was also observed \cite{Dowker:1993bt,Dowker:1995gb} that for the particular value of dilaton coupling corresponding to Kaluza-Klein reduction from $5D$ vacuum gravity, the dilatonic Melvin solution \cite{Gibbons:1987ps} arises from reduction of flat $5D$ Minkowski spacetime with certain twisted identifications.  A string theory interpretation of the results of this construction in Kaluza-Klein reduction from $D=11$ supergravity to $D=10$ Type $IIA$ supergravity was given in \cite{Costa:2000nw}.  The general notion of flux $p$-branes, Melvin-like solutions that are translation invariant in $p$ spatial directions and time, was developed in \cite{Gutperle:2001mb} together with a number of further examples, including a supersymmetric flux $5$-brane.

Our focus in this paper is on properties of Melvin fluxtubes in anti-deSitter backgrounds, which have been found in \cite{Astorino:2012zm} (see also \cite{Lim:2018vbq}).  We find a number of interesting new features.  
For $\Lambda=0$ Melvin fluxtubes are not asymptotically flat.   As the radial coordinate is increased beyond the fluxtube radius, circles surrounding the fluxtube begin to decrease in circumference, with the circumference shrinking to zero at infinite radius. 
One cannot then {\it e.g.} compute the mass per unit length of a $\Lambda=0$ Melvin fluxtube.  AdS-Melvin spacetimes, however, are asymptotically AdS and we can compute their ADM charges.  

We also find consequences of the confining, box-like properties of AdS.  The characteristic radius of $\Lambda=0$ Melvin fluxtubes goes inversely with the strength\footnote{We will assume in this paper that $B\ge 0$ and hence also that all magnetic fluxes are positive.   This will simplify a number of formulas which would otherwise require absolute value signs.}  $B$ of the magnetic field.    $\Lambda=0$ fluxtubes with strong magnetic fields are narrow and carry only small quantities of magnetic flux, while fluxtubes with weak magnetic fields are broad and carry large amounts of magnetic flux.  The total flux diverges as the limit of vanishing field strength is approached, since a field of strength $B$ covers an area that scales as $1/B^2$.   In AdS, the behavior of fluxtubes with strong magnetic fields and radii much smaller than the AdS length scale is essentially unchanged.  However, as $B$ is decreased the radii of fluxtubes are now cutoff at a scale of order the AdS length scale.  The total magnetic flux of an AdS-Melvin spacetimes then vanishes as $B\rightarrow 0$, since a smaller and smaller field strength is now spread out over only a finite area.

Since the total flux of an AdS-Melvin fluxtube vanishes both in the limits of small and large $B$, it must have a maximum for some intermediate value of $B$.  We find that the maximum magnetic flux that can be accommodated in AdS in the form of a static AdS-Melvin fluxtube is given by $\phimax =2\pi/\sqrt{-\Lambda}$.  Although it is possible that other static flux configurations exist, this suggests that spacetimes with flux $\phitot >\phimax$ will be dynamical, and given the confining nature of AdS it seems likely that the fate of such configurations is gravitational collapse.  On the other hand, for $\phitot <\phimax$ there are two AdS-Melvin fluxtubes with different values of the magnetic field strength $B$.  In the limit of small flux, $\phitot\ll\phimax$, one of these fluxtubes will be narrow, while the other will have radius of order the AdS length scale.  It seems likely that one of these branches of solutions will be stable and the other unstable.  We will not carry out a stability analysis here.  However, we will offer a number of comments regarding the possible (in)stability of the two branches.  In particular, we find that the broader branch of fluxtubes always has lower mass and connects smoothly for $B=0$ to the AdS soliton \cite{Horowitz:1998ha}. 

The paper proceeds in the following way.  In section (\ref{melvinsection}) we recall basic properties of the Melvin spacetime.  In section (\ref{adsmelvinsection}) we present the AdS-Melvin spacetimes \cite{Astorino:2012zm} and explore the properties noted above.  In section (\ref{ADMsection}) we compute the ADM mass and tensions of AdS-Melvin spacetimes.  Finally we offer some conclusions and notes on directions for future work in section (\ref{conclusions}).

\section{Melvin Spacetime}\label{melvinsection}
We will be interested in solutions to $4D$ Einstein-Maxwell theory with a cosmological constant
\begin{equation}\label{action}
S = \int d^4x\sqrt{-g}\left(R-2\Lambda-F^2\right)
\end{equation}
and particularly in the case $\Lambda<0$.  However, we begin by recalling properties of the Melvin solution \cite{Melvin:1963qx} with $\Lambda=0$,
which describes a static, cylindrically symmetric bundle of magnetic flux
\begin{align}\label{melvin}
ds^2 &= (1+{B^2\rho^2\over 4})^2\ \left( -dt^2 + dx^2 +d\rho^2\right) +{1\over (1+{B^2\rho^2\over 4})^2}\rho^2d\phi^2\\
A_\phi & = {2\over B}\left({{B^2\rho^2\over 4}\over  1+{B^2\rho^2\over 4}}\right)   \nonumber
\end{align}
In addition to its invariance under translations in $t$, $x$, the spacetime has a boost symmetry in the $x$-direction.  The $\phi$ coordinate is usually assumed to be identified $\phi\equiv\phi +2\pi$, so that translation symmetry in $\phi$ can be thought of as rotation symmetry around the axis at $\rho=0$.
Geometric properties of the spacetime are characterized by the Melvin radius
\begin{equation}\label{magnetlength}
\rho_M= {2\over |B|}
\end{equation}
Near the rotational axis, {\it i.e.} for $\rho\ll \rho_M$, the metric is approximately that of flat spacetime in cylindrical coordinates.    
The Melvin spacetime is not asymptotically flat at large $\rho$, instead having the asymptotic form for $\rho\gg\rho_M$
\begin{equation}\label{levicivita}
ds^2\simeq \left({\rho\over\rho_M}\right )^4\left( -dt^2 + dx^2 +d\rho^2\right)+\left({\rho_M^4\over \rho^2}\right)d\phi^2
\end{equation}
This limiting form coincides with a non-flat solution of the static, cylindrically symmetric, vacuum Levi-Civita class (see reference \cite{Kastor:2015wda} for a discussion).  

\begin{figure}[h]
\begin{center}\includegraphics[width=0.50\textwidth]{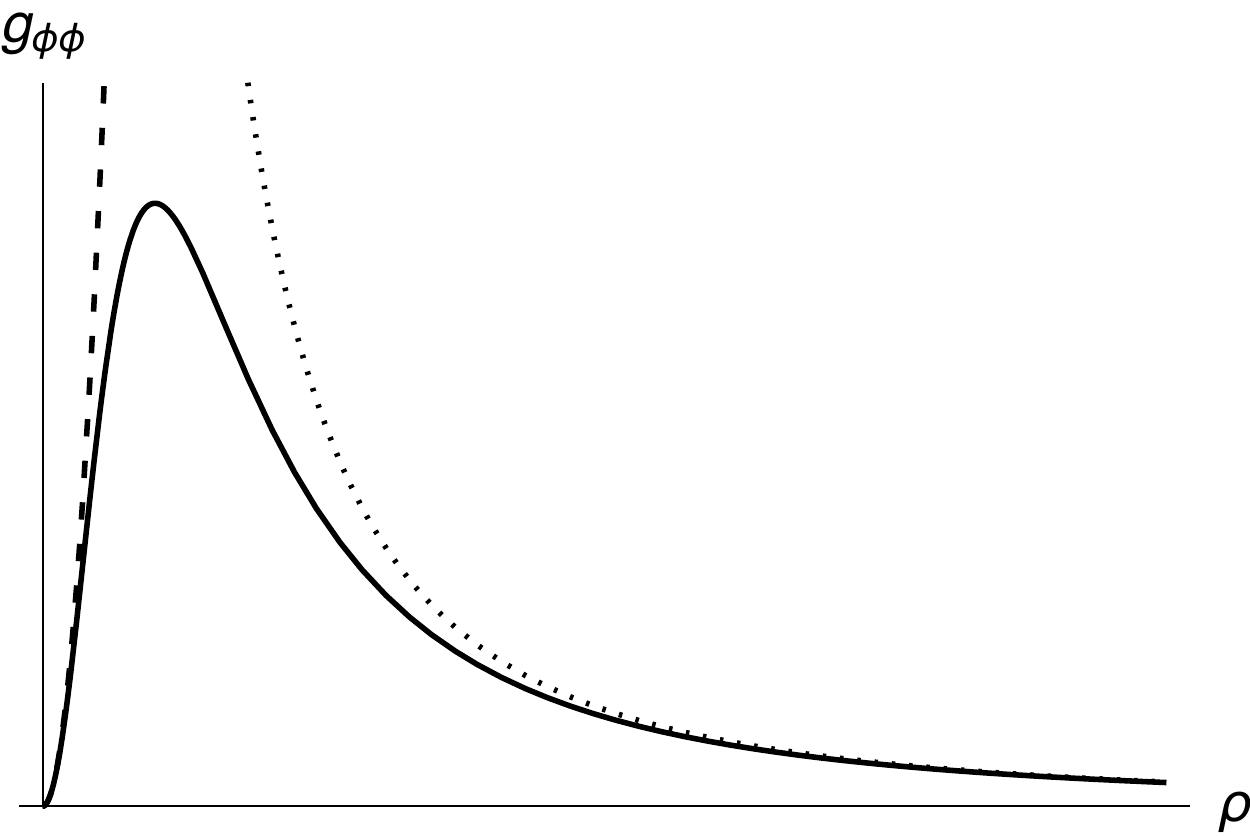}\end{center}
\caption{\it Plot of the angular metric component $g_{\phi\phi}$ versus radius for the Melvin spacetime.}  
\label{melvinfig}
\end{figure}

The behavior of the metric component $g_{\phi\phi}$ for the Melvin spacetime (\ref{melvin}) is shown by the solid line in figure (\ref{melvinfig}).  For small radii, such that $\rho\ll\rho_M$, the angular component of the metric tracks that of flat spacetime in cylindrical coordinates, shown by the dashed line, while asymptotically for large radii it comes into agreement with non-flat Levi-Civita form in (\ref{levicivita}) shown by the dotted line.
Correspondingly, the circumferences of circles of sufficiently small constant values of $\rho$ initially increase with $\rho$ as in flat spacetime.  However, for larger values of $\rho$ the circumference reaches a maximum value before decreasing asymptotically to zero size.

The profile of the magnetic field in the Melvin spacetime can be understood by considering the magnetic flux $\Phi(\rho)$ contained within a circle of radius $\rho$.  The flux is given simply by the line integral of the gauge potential 
\begin{align}\label{flux}
\Phi(\rho) &=  \oint A_\phi (\rho) d\phi \\ &={4\pi\over B}{{B^2\rho^2\over 4}\over  1+{B^2\rho^2\over 4}}
\end{align}
This has the following limiting behavior for circles either much smaller, or much larger, than the Melvin radius
\begin{equation}
\Phi(\rho) \simeq\left\{ 
\begin{array}{lr}
B\pi\rho^2, & \rho\ll\rho_M\\
{4\pi\over B}, & \rho\gg\rho_M
\end{array}\right .
\end{equation}
For small circles, the flux grows like a constant magnetic field strength $B$ times the flat space area $\pi\rho^2$ for a circle of radius $\rho$.   However, for circles larger than the Melvin radius the flux approaches the asymptotic value
\begin{equation}\label{flux}
\phitot
= {4\pi\over B}
\end{equation}
The majority of the magnetic flux is contained within a few Melvin radii surrounding the $\rho=0$ axis and it is reasonable to think of $\rho_M=2/|B|$ as an approximate fluxtube radius.  Due to the inverse relationship between the fluxtube radius and the field strength, fluxtubes with strong field strengths are narrow and carry little total flux $\Phi$.  Fluxtubes carrying weak magnetic fields, by contrast, have such a large cross section that the flux $\Phi$ tends to infinity in the limit of weak fields.  We will see that this latter property is dramatically changed in AdS.

\section{AdS-Melvin spacetimes}\label{adsmelvinsection}

An AdS generalization of Melvin was found in \cite{Astorino:2012zm} (see also \cite{Lim:2018vbq}).  However, its physical properties have not been extensively explored.  The construction in \cite{Astorino:2012zm} employs a generalization to $\Lambda\neq 0$ of the solution generating tool of \cite{Ernst:1976mzr}.  With $\Lambda=0$, this technique produces the Melvin spacetime (\ref{melvin}) starting from flat spacetime in cylindrical coordinates as a seed metric.  The seed metric used in  \cite{Astorino:2012zm} is not the AdS metric, but rather the AdS soliton metric \cite{Horowitz:1998ha}, which shares certain important features with the flat metric in cylindrical coordinates\footnote{Note that the AdS soliton spacetime has negative mass with respect to the AdS asymptotics \cite{Horowitz:1998ha}.  It is conjectured in \cite{Horowitz:1998ha} that the soliton is the lowest energy state with a compact spatial coordinate and also  in this sense, it seems like a natural starting point for the AdS-Melvin construction.}.  Because the AdS-Melvin spacetime will reduce to the AdS soliton for $B=0$, we start by recalling its basic features.


The AdS soliton \cite{Horowitz:1998ha} is obtained via analytic continuation from the planar AdS black hole, and the metric is given by
\begin{equation}\label{adssoliton}
ds^2 = {l^2\over r^2}{dr^2\over F(r)} +{r^2\over l^2}\left( -dt^2 +dx^2+F(r)dy^2\right),\qquad
F(r) = 1-{r_0^3\over r^3}
\end{equation}
where $l$ is the AdS length scale given by $l^2= -3/\Lambda$.
The soliton radius $r_s$ defined by the condition $F(r_s)=0$ is given simply by $r_s=r_0$.  The $y$-coordinate is assumed to be periodically identified
\begin{equation}
y\equiv y+L_y
\end{equation}
with periodicity chosen such that the spacetime caps off smoothly at the soliton radius, in the manner of a Euclidean black hole.  The period $L_y$ is the analogue of the inverse black hole temperature.  To identify the correct period $L_y$, we can define new radial and angular coordinates $(\rho,\phi)$ by means of the coordinate transformation
\begin{equation}\label{coordtrans}
r = (1+{3\rho^2\over 4l^2})\, r_s,\qquad y ={2l^2\over 3r_s}\phi
\end{equation}
so that the soliton is now located at $\rho=0$.
The AdS soliton metric is now given in these new coordinates by
\begin{align}\label{adssoliton2}
ds^2 = &{(1+{3\rho^2\over 4l^2})\over H(\rho)}d\rho^2 + {H(\rho)\over (1+{3\rho^2\over 4l^2})}\rho^2 d\phi^2 +{r_0^2\over l^2}(1+{3\rho^2\over 4l^2})^2(-dt^2 +dx^2)\\
H(\rho)=&1+{3\rho^2\over 4 l^2}+{3\rho^4\over 16 l^4}\nonumber
\end{align}
Focusing in on a small patch around the soliton radius with $\rho/l\ll 1$, we see that the metric is approximately the flat metric in cylindrical coordinates
\begin{equation}
ds^2\simeq d\rho^2 +\rho^2 d\phi^2 + {r_0^2\over l^2}(-dt^2 +dx^2) 
\end{equation}
This will be smooth at $\rho=0$ if the angular coordinate $\phi$ is identified with period $2\pi$.  We then see from (\ref{coordtrans}) that the correct identification $L_y$ for the original $y$ coordinate is given by 
\begin{equation}\label{yperiod}
L_y = {4\pi l^2\over 3 r_s}
\end{equation}
Recalling that the flat seed metric that yielded the $\Lambda=0$ Melvin spacetime via solution generating \cite{Ernst:1976mzr} was the  Minkowski metric in cylindrical coordinates, the fact that the AdS soliton has this same limiting form for small $\rho$ makes it the appropriate starting point for the generalized solution generating method of \cite{Astorino:2012zm}.


It was pointed out \cite{Lim:2018vbq} that the AdS-Melvin solution obtained by solution generating in \cite{Astorino:2012zm} is equivalent under a coordinate transformation to a magnetized version of the AdS soliton given by
\begin{align}\label{adsmelvin}
ds^2 =& {l^2\over r^2}{dr^2\over F(r)} +{r^2\over l^2}\left( -dt^2 +dx^2+F(r)dy^2\right), \\
F(r) = &1-{r_0^3\over r^3}-{B^2l^6\over r^4},\qquad A_y=-{Bl^2\over r}\nonumber
\end{align}
which is itself an analytic continuation of the charged planar AdS black hole.   We want to understand what properties this magnetized soliton spacetime shares with $\Lambda=0$ Melvin (\ref{melvin}) and what aspects are qualitatively changed by the presence of a nonzero, negative cosmological constant.

A first observation is simply that the AdS-Melvin spacetime (\ref{adsmelvin}) is asymptotically AdS.  The new ${\mathcal O}(1/r^4)$ term added to the metric function $F(r)$ does not change the leading asymptotic behavior relative to the AdS soliton spacetime. 
This contrasts with $\Lambda=0$ Melvin, which fails to be asymptotically flat.  However, it is consistent with the expectation that AdS generally acts like a box, in this case confining the Melvin magnetic flux\footnote{It also appears consistent with the $\Lambda>0$ results   of \cite{Wald:1983ky}, in which the effects of a positive cosmological constant dominate at infinity over those of matter satisfying a reasonable energy condition, although we have not analyzed this in detail for $\Lambda<0$.}.  These confining properties of AdS effectively regulate the poor asymptotic behavior of $\Lambda=0$ Melvin.  In particular, this will allow one to compute ADM charges for AdS-Melvin spacetimes (see \cite{El-Menoufi:2013pza} for the relevant formulas) and to analyze the thermodynamics of AdS-Melvin spacetimes.

While AdS dominates at large radius, the region surrounding the soliton exhibits Melvin-type physics.  For a fixed value of $B$, we organize the analysis by specifying the soliton radius $r_s>0$ defined by $F(r_s)=0$.  The parameter $r_0$ in the metric (\ref{adsmelvin}) is then given by
\begin{equation}\label{rzero}
r_0^3 = (1-\bhat^2 l^2)r_s^3
\end{equation}
Here $\bhat$ is a rescaled magnetic field strength parameter given by\footnote{The AdS-Melvin spacetime presented in \cite{Astorino:2012zm} is restricted to the case $r_s=l$. }
\begin{equation}\label{bhat}
\bhat = {l^2\over r_s^2}B
\end{equation}
Again we will require that the spacetime cap off smoothly at the soliton radius.
In order to identify the correct periodicity $L_y$, we transform to the new radial and angular coordinates $(\rho,\phi)$ given by the coordinate transformation
\begin{align}
r & = (1+\alpha\rho^2)r_s,\qquad y ={1\over 2\alpha r_s}\phi
\end{align}
where the parameter $\alpha$ is given by
\begin{equation}
\alpha  = {\bhat^2\over 4}(1+{3\over \bhat^2 l^2})  \label{alpha}
\end{equation}
With $B=0$ these expressions reduce to the coordinate transformations (\ref{coordtrans}) used above for the unmagnetized AdS soliton.  
The AdS-Melvin spacetime now has the form in these new coordinates
\begin{align}\label{adsmelvin2}
ds^2 &= (1+\alpha\rho^2)^2\left (-d\that^2 +d\xhat^2 +{d\rho^2\over H(\rho)}\right)
+ {H(\rho)\over (1+\alpha\rho^2)^{2}} \rho^2d\phi^2   \\
A_\phi& = {2\over \bhat(1+{3\over \bhat^2 l^2})}\left({\alpha\rho^2\over 1+\alpha\rho^2}\right),\qquad 
H(\rho)= 1 +{3\rho^2\over 2l^2}(1+{2\over 3}\alpha\rho^2 +{1\over 6}\alpha^2\rho^4)
\end{align}
where $\that=r_st/l$ and $\xhat = r_s x/l$ and a gauge transformation has been made such that $A_\phi$ vanishes at $\rho=0$.  
The metric will be regular at $\rho=0$ if the angular coordinate $\phi$ has the standard identification 
$\phi\equiv\phi+2\pi$.   This implies that the periodicity of the original $y$-coordinate periodicity should be taken to be
\begin{equation}\label{Ly}
L_y = {4\pi\over r_s\bhat^2(1+{3\over \bhat^2 l^2})}
\end{equation}
From a physical perspective, it makes sense to choose $L_y$ as one of the independent parameters specifying a solution.  

We can now compare AdS-Melvin spacetime in the form (\ref{adsmelvin2}) with the original $\Lambda=0$  Melvin spacetime (\ref{melvin}).    As noted in \cite{Astorino:2012zm}, AdS-Melvin reduces correctly to $\Lambda=0$ Melvin when the AdS length scale $l$ is taken to infinity with $\bhat$ held fixed.  We can also see a Melvin regime emerge in (\ref{adsmelvin2}) for finite $l$, by restricting our attention to the regime $\rho\ll l$ for fluxtubes that are narrow compared to the AdS radius, which we will see below is the condition $\bhat^2 l^2\gg 1$.  In this regime the parameter $\alpha\simeq \bhat^2/4$ and the function $H(\rho)\simeq 1$, giving the Melvin form of the metric (\ref{melvin}) to close approximation.

\begin{figure}[ht]
\centering
\subfigure[Narrow fluxtube]{\includegraphics[width=0.40\textwidth]{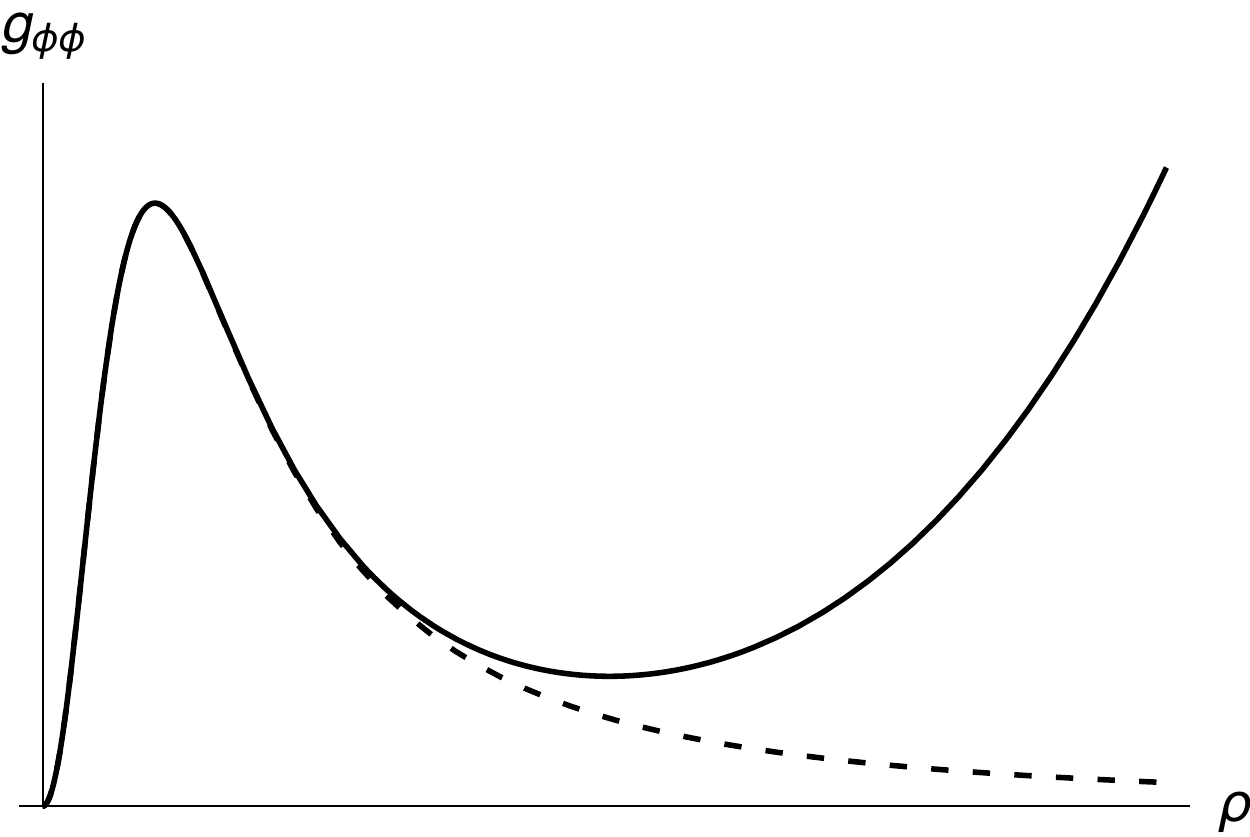}  \label{narrow}}
\qquad
\subfigure[Broad fluxtube]{\includegraphics[width=0.40\textwidth]{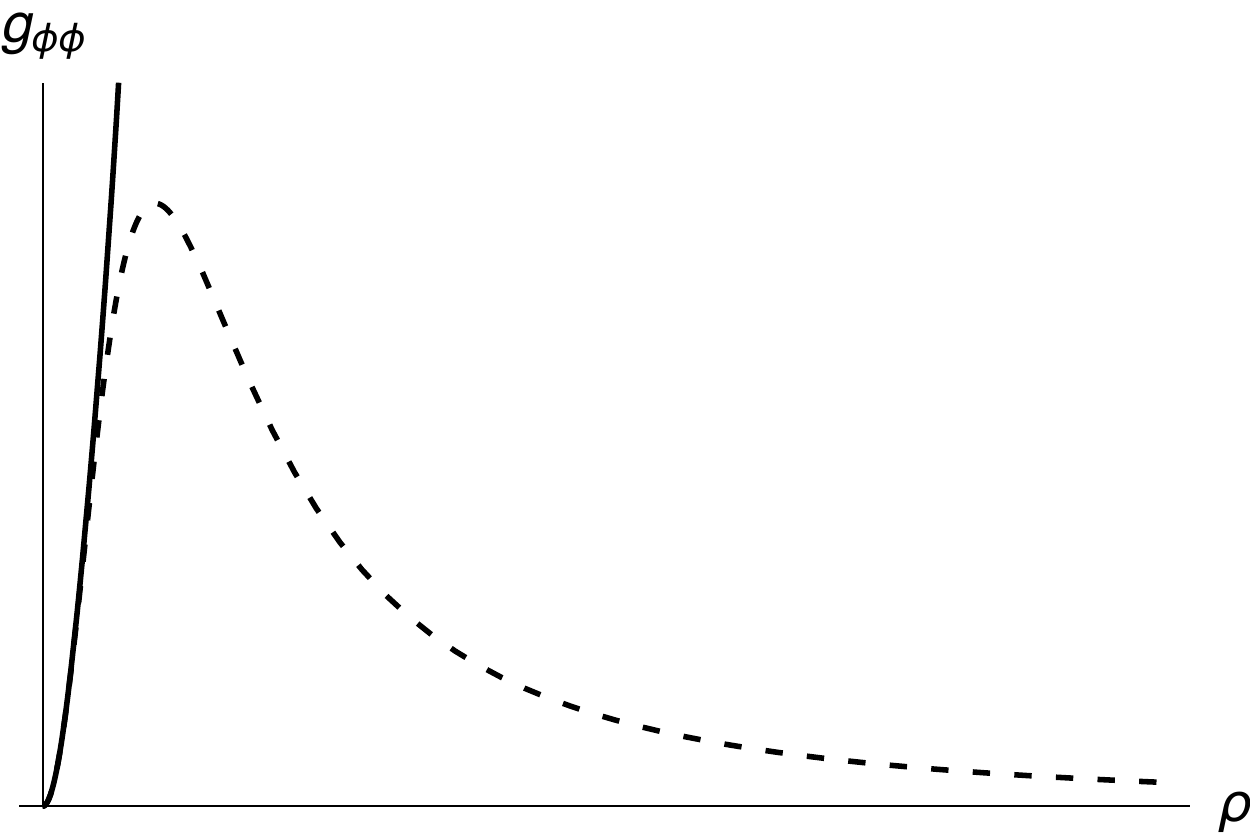}\label{broad}}
\caption{\it Plots comparing the radial evolutions of narrow and broad AdS-Melvin fluxtubes in comparison with the form of Melvin fluxtubes.  Note that the horizontal axis on the right has been rescaled in order to facilitate this comparison.  The rescaling causes the broad fluxtube to appear narrower than its narrow counterpart on the left.}
\label{fig:figure}
\end{figure}

The coincidence of narrow AdS-Melvin fluxtubes with Melvin is illustrated in figure (\ref{narrow}), which shows the metric component $g_{\phi\phi}$ as a function of the radial coordinate $\rho$ for a fluxtube with $\bhat l=200$.  The dashed line shows the Melvin form of $g_{\phi\phi}$, while the solid line shows the AdS-Melvin behavior.  The AdS-Melvin fluxtube tracks the Melvin form through the peak and subsequent falloff in the size of circles of constant $\rho$ before departing to its asymptotically AdS form.  Increasing $\bhat l$ further would expand the region through which AdS-Melvin tracks the $\Lambda=0$ Melvin form.  A broader fluxtube with $\bhat l=2$ shown in figure (\ref{broad}) departs from the Melvin form at a much earlier stage of its radial development.  The circumference of circles in broad AdS-Melvin fluxtubes increase monotonically with radius.


The magnetic flux $\Phi(\rho)$ enclosed by a circle of radius $\rho$ in the AdS-Melvin spacetime is computed from the line integral around the circle of the gauge potential in (\ref{adsmelvin2}) with the result that
\begin{equation}\label{adsflux}
\Phi(\rho) = {\alpha\rho^2\over 1+\alpha\rho^2}\, \phitot,\qquad \phitot = {4\pi\over \bhat(1+{3\over \bhat^2 l^2})}
\end{equation}
where $\phitot$ is the total magnetic flux of the spacetime.  The radial profile of the flux is characterized by the AdS-Melvin radius
\begin{equation}\label{adsmelvinradius}
\rho_{M} = {1\over\sqrt{\alpha}} = {2\over \bhat\sqrt{1+{3\over \bhat^2 l^2}}}
\end{equation}
The behavior of the AdS-Melvin radius is controlled by the dimensionless parameter $\bhat^2 l^2$, such that
\begin{equation}\label{melvinlengthlimits}
\rho_M \simeq\left\{ 
\begin{array}{lr}
{2\over \bhat} & \bhat^2 l^2\gg 1\\
{2l\over\sqrt{3}}, & \bhat^2 l^2\ll 1
\end{array}\right .
\end{equation}
We see that fluxtubes with large field strength $\bhat$, such that $\bhat^2 l^2\gg 1$, are indeed  narrow in correspondence with the $\Lambda=0$ Melvin radius (\ref{magnetlength}).  However, the behavior of $\rho_M$ is qualitatively modified by $\Lambda<0$ at weak field strength, such that $\bhat^2 l^2\ll 1$.  In this limit the AdS-Melvin radius is comparable to the AdS radius, whereas the $\Lambda=0$ Melvin radius diverges.  AdS spacetime is acting as a confining box for the Melvin magnetic flux.

The corresponding behavior for the total flux $\phitot$  in these regimes is
\begin{equation}
\phitot  \simeq\left\{ 
\begin{array}{cc}
{4\pi\over \bhat}, & \bhat^2l^2\gg 1\\
\pi\rho_M^2\bhat, &  \bhat^2l^2\ll 1
\end{array}\right .
\end{equation}
Again this coincides with the result for $\Lambda=0$ Melvin for fluxtubes that are much narrower than the AdS length scale, such that $\bhat^2l^2\gg 1$.  Although the field strength inside the fluxtube is large in this regime, the fluxtube radius is going to zero sufficiently fast that the total flux goes to zero as the limit is approached.  Turning to the weak field regime, such that $\bhat^2l^2\ll 1$, we see that the flux $\phitot$ also approaches zero in this limit, since a decreasing magnetic field strength is now spread over an area of only finite radius.  This is in contrast to the $\Lambda=0$ case in which the total flux diverges, weak field strength being spread throughout an increasingly large region.  The vanishing, rather than divergence, of $\phitot$ as $\bhat$ tends to zero is a further aspect of how Melvin fluxtubes are regulated by the presence of an AdS background.


$\Lambda=0$ Melvin has a simple relationship (\ref{flux}) between the field strength and total magnetic flux, such that the total flux decreases monotonically to zero as $B$ increases.
The flux $\phitot$ of AdS-Melvin (\ref{adsflux}) spacetimes, however, vanishes in both the limits of small and large field strengths $\bhat$, and therefore must have a maximum value $\phimax$ for some intermediate value of $\bhat$. This is found to be
\begin{equation}
\phimax = {2\pi l\over\sqrt{3}} = {2\pi\over\sqrt{-\Lambda}}
\end{equation}
and occurs for magnetic field strength $\bhat_{max}=\sqrt{-\Lambda}$.   We see then that AdS can only accommodate a limited quantity of magnetic flux in a static Melvin-like configuration.  Note that this maximum value correctly diverges as $\Lambda$ tends to zero.

For all smaller values of the flux, there will be two corresponding magnetic fluxtubes with magnetic field strengths
\begin{equation}\label{bhatpm}
\bhat_\pm = {2\pi\over \phitot}\left(1\pm\sqrt{1-{\phitot^2\over\phimax^2}}\,\right)
\end{equation}
Looking at $\phitot\ll\phimax$, we see that $\bhat_+$ diverges in this limit corresponding to narrow fluxtubes. This branch of solutions matches on to $\Lambda=0$ Melvin as the cosmological constant is taken to zero.  For the $\bhat_-$ branch, the magnetic field strength goes continuously to zero with the flux and corresponds to flux tubes with widths of order the AdS length scale.  This branch will match continuously onto the AdS soliton metric as $\Phi$ and $\bhat_-$ tends to zero.  

With two branches of solutions, it is often the case that one branch is stable and the other unstable.  In this case, both branches have {\it a priori} claims to stability.   $\Lambda=0$ Melvin fluxtubes have been found to be stable \cite{Melvin:1965zza,thorne_stability}.   So, it is natural to conjecture that the $\bhat_+$ branch of solutions with $\Lambda<0$ will continue to be stable.  Given that AdS typically acts as a confining box, we would also expect that the more compact flux configuration would tend to be stable.  On the other hand, the $\bhat_-$ branch is continuously connected to the AdS soliton, which is conjectured \cite{Horowitz:1998ha} to be the lowest energy state with fixed periodicity $L_y$ in the $y$-direction (see {\it e.g.} \cite{Woolgar:2016axs,Barzegar:2019vaj} for recent work in this area).  This might lead one to expect that the $\bhat_-$ branch should continue to be stable.  We will not carry out a stability analysis here, but we will show that if we fix the periodicity $L_y$ and the total magnetic flux $\phitot$, then the $\bhat_-$ branch always has lower mass.


\section{ADM mass and tensions}\label{ADMsection}

$\Lambda=0$ Melvin spacetimes are not asymptotically flat, and therefore one cannot compute {\it e.g.} the mass of a Melvin  spacetime relative to a flat background.   However, AdS-Melvin spacetimes are asymptotically planar AdS and so one can compute their ADM charges.  The list of charges includes, in addition to the mass $\calm$, tensions $\calt_x$ and $\calt_y$ in the $x$ and $y$ directions  (see \cite{El-Menoufi:2013pza,Kastor:2018cqc}).  Tension charges are defined similarly to the mass, but with respect to asymptotic spatial translation Killing vectors and contribute to stress-strain terms in the first law when the length of a compact direction is varied \cite{Kastor:2006ti}.

An asymptotically planar AdS spacetime in $4$-dimensions has the following form in the limit of large radial coordinate $r$
\begin{equation}
ds^2\simeq  {l^2\over r^2} (1+{c_r\over r^3})dr^2 +  {r^2\over l^2}\left( (-1+{c_t\over r^3})dt^2 + (1+{c_x\over r^3})dx^2 + (1+{c_y\over r^3})dy^2\right)
\end{equation}
where by virtue of the field equations the fall-off coefficients satisfy the sum rule $c_r-c_t+c_x+c_y = 0$.
In order that the ADM charges be finite, the spatial coordinates $x$ and $y$ are assumed to be identified with periods
\begin{equation}
x\equiv x+L_x,\qquad y\equiv y+L_y
\end{equation}
and the two dimensional volume $v$ is defined by $v=L_xL_y$.
The ADM mass and tensions are then given by
\begin{align}\nonumber
\calm &= {v\over 16\pi l^4}(2c_r+3(c_x+c_y))\\
\calt_x L_x& = {v\over 16\pi l^4}(2c_r+3(-c_t+c_y))\\
\calt_y L_y& = {v\over 16\pi l^4}(2c_r+3(-c_t+c_x))\nonumber
\end{align}
The ADM mass and tensions satisfy the sum rule\footnote{The factors of $L_x$ and $L_y$ occurring along with the corresponding tensions arise because the boundary integrals defining $\calt_x$ and $\calt_y$ do not include integraling respectively over the $x$ and $y$ directions.  This is analogous to the integral for the mass not including the time direction.  Strictly speaking the tension charges are tensions per unit time, which accounts for their different dimensionality from the mass \cite{Kastor:2006ti}.}
$\calm+\calt_xL_x+\calt_yL_y = 0$ as a consequence of the scaling Killing vector of AdS \cite{El-Menoufi:2013tca}.  The sum rule is closely tied to conformal invariance of the boundary theory and tracelessness of the boundary stress tensor.

We start by evaluating the ADM charges for the original AdS soliton spacetime (\ref{adssoliton}), fixing the period of identification $L_y$ of the $y$-coordinate.  The falloff coefficients are seen to be $c_r=-c_y=r_s^3$ and $c_t=c_x=0$, with the soliton radius determined in terms of $L_y$ by (\ref{yperiod}).  The mass and tensions for the AdS soliton are then found to be 
\begin{equation}\label{solitonmass}
\calm = \calt_x L_x  = - {4\pi^2 l^2 L_x\over 27 L_y^2}, \qquad \calt_y L_y =+ {8\pi^2 l^2 L_x\over 27 L_y^2}
\end{equation}
%
in agreement with \cite{Horowitz:1998ha}.  The negative soliton mass is attributed in the context of holography to the Casimir energy of the CFT confined to the finite $S^1$ in the $y$-direction \cite{Horowitz:1998ha}.
It is further conjectured that the AdS soliton is indeed the lowest mass state with these boundary conditions \cite{Horowitz:1998ha}.  If the AdS soliton is the ground state with these boundary conditions, then its role (analogous to flat spacetime in the $\Lambda=0$ case) as the seed metric for the construction of AdS-Melvin spacetimes \cite{Astorino:2012zm} seems natural.

We now compute the ADM mass and tensions for AdS-Melvin spacetimes.   There are different choices that one can make for the independent variables.  We adopt a physical perspective and fix the periodicity $L_y$ and the total magnetic flux $\phitot$.  Given $L_y$ and $\phitot\le\phimax$, there are then two solutions with different values of $B$ and $r_s$, such that the rescaled field strength parameter takes either of the values $\bhat_\pm$ given in (\ref{bhatpm}).

The falloff coefficients are then given by
\begin{equation}
c_r = -c_y= (1-\bhat^2_\pm l^2)r_s^3,\qquad c_t=c_x=0
\end{equation}
where we have made use of equation (\ref{rzero}) which gives the parameter $r_0$ in the AdS-Melvin metric  in terms of the soliton radius $r_s$ and the magnetic field strength parameter $\bhat$.  These latter quantities are found to be given in terms of $L_y$ and $\phitot$ for the two branches of solutions by 
\begin{equation}
r_{s\pm}={2\pi l^2\over 3L_y}\left(1\mp\sqrt{1-{\phitot^2\over\phimax^2}}\right),\qquad 
B_\pm = {2\pi \phitot\over 3 L_y^2}\left(1\mp\sqrt{1-{\phitot^2\over\phimax^2}}\right)
\end{equation}
The masses of the two solutions are then found to be
\begin{equation}
\calm_\pm = -{2\pi l^2 L_x\over 27 L_y^2}\left\{1 -{3\phitot^2\over 2\phimax^2}\mp \left(1 - {\phitot^2\over \phimax^2}\right)^{{3\over 2}}\right\}
\end{equation}
while the tensions are given by $\calt_xL_x=\calm$ and $\calt_yL_y=-2\calm$.  Recalling that the `-' branch of solutions matches smoothly to the AdS soliton as the flux is taken to zero, we see that $\calm_-$ correctly reproduces the AdS soliton mass (\ref{solitonmass}) in this limit.  Moreover, we see that the mass splitting between the two branches is given by
\begin{equation}
\calm_+-\calm_- = -\calm_s \left(1 - {\phitot^2\over \phimax^2}\right)^{{3\over 2}}\ge 0
\end{equation}
where $\calm_s$ is the AdS soliton mass (\ref{solitonmass}).  This shows that the narrower AdS-Melvin spacetimes on `+' branch of solutions always have greater mass than the fluxtubes on the `-' branch that have radius of order the AdS length scale.  We can note also that $\calm_\pm\ge \calm_s$ in accordance with the positive mass conjecture of \cite{Horowitz:1998ha}.


\section{Conclusions}\label{conclusions}

We have studied geometric features of AdS-Melvin spacetimes \cite{Astorino:2012zm} and found a number of interesting distinctions from the behavior of $\Lambda=0$ Melvin spacetimes.  We have seen that these spacetimes display the confining properties of AdS, with the fluxtube radius cut off by the AdS radius.  This is qualitatively different from the behavior of 
$\Lambda=0$ fluxtubes which can have arbitrarily large radii.  As a result of this cut off, there is a maximum magnetic flux $\phimax=2\pi/\sqrt{-\Lambda}$ for Melvin fluxtubes in AdS.  For $|\Phi|<\phimax$, there are two branches of fluxtubes with the same periodic identification $L_y$.  From our computation of the ADM charges we have seen that the branch of solutions with wider radii and lower field strength always has the lower mass.

We comment on a number of possible directions for future work.  It would be interesting to probe the stability of the two branches of AdS-Melvin fluxtubes with degenerate flux and $L_y$.  In particular, it would be nice to know which, if any, of the reasons for (in)stability given above are valid.  It would likewise be interesting to explore the thermodynamics of AdS-Melvin spacetimes.  In this context, it is known that the surface area of the soliton will make an entropy-like contribution \cite{El-Menoufi:2013pza,Kastor:2008wd}.  Generalized Melvin spacetimes have been found that interpolate between different vacuum Levi-Civita regimes near the axis and near infinity \cite{Kastor:2015wda}.  AdS analogues of Levi-Civita spacetimes are also known \cite{Tian:1986zz,Linet:1986sr} (see also \cite{Kastor:2018cqc}).  Work is in progress \cite{AKST} to see whether generalized AdS-Melvin spacetimes can also be constructed by means of solution generating methods \cite{Astorino:2012zm}.

Finally, it would be interesting to study deSitter analogues of the spacetimes considered here.  Cosmological analogues of Melvin spacetimes may be obtained from (\ref{melvin}) by simple analytic continuation \cite{Kastor:2015wda} (see \cite{Sabra:2020gio} for higher dimensional generalizations).  Similarly, the AdS soliton solution analytically continues to a member of the dS-Kasner family of spacetimes (see \cite{Kastor:2016bnm}), which have been studied as anisotropic models of inflation (see {\it e.g.} \cite{Gumrukcuoglu:2007bx,Blanco-Pillado:2015dfa}).  Analytic continuation of AdS-Melvin spacetimes and their generalizations \cite{AKST} will give anisotropic inflationary cosmologies with early electromagnetic phases.

\section*{Acknowledgements} We thank Marco Astorino and Wafic Sabra for discussions and for ongoing collaboration on related issues.


\begin{thebibliography}{99}

\bibitem{Melvin:1963qx}
M.~Melvin,
``Pure magnetic and electric geons,''
Phys. Lett. \textbf{8}, 65-70 (1964)
doi:10.1016/0031-9163(64)90801-7

\bibitem{Melvin:1965zza}
M.~A.~Melvin,
``Dynamics of Cylindrical Electromagnetic Universes,''
Phys. Rev. \textbf{139}, B225-B243 (1965)
doi:10.1103/PhysRev.139.B225

\bibitem{thorne_stability}
K.~S.~Thorne,
``Absolute Stability of Melvin's Magnetic Universe,"
Phys. Rev. \textbf{139}, B244-B254 (1965)

\bibitem{Ernst:1976mzr}
F.~J.~Ernst,
``Black holes in a magnetic universe,''
J. Math. Phys. \textbf{17}, 54 (1976)
doi:10.1063/1.522781

\bibitem{Gibbons:2013yq}
G.~W.~Gibbons, A.~H.~Mujtaba and C.~N.~Pope,
``Ergoregions in Magnetised Black Hole Spacetimes,''
Class. Quant. Grav. \textbf{30}, no.12, 125008 (2013)
doi:10.1088/0264-9381/30/12/125008
[arXiv:1301.3927 [gr-qc]].

\bibitem{Gibbons:2013dna}
G.~W.~Gibbons, Y.~Pang and C.~N.~Pope,
``Thermodynamics of magnetized Kerr-Newman black holes,''
Phys. Rev. D \textbf{89}, no.4, 044029 (2014)
doi:10.1103/PhysRevD.89.044029
[arXiv:1310.3286 [hep-th]].

\bibitem{Cvetic:2013roa}
M.~Cvetic, G.~W.~Gibbons, C.~N.~Pope and Z.~H.~Saleem,
``Electrodynamics of Black Holes in STU Supergravity,''
JHEP \textbf{09}, 001 (2014)
doi:10.1007/JHEP09(2014)001
[arXiv:1310.5717 [hep-th]].


\bibitem{Kinnersley:1970zw}
W.~Kinnersley and M.~Walker,
``Uniformly accelerating charged mass in general relativity,''
Phys. Rev. D \textbf{2}, 1359-1370 (1970)
doi:10.1103/PhysRevD.2.1359

\bibitem{ernst-cmetric}
F.~J.~Ernst,
``Removal of the nodal singularity of the C?metric,''
J. Math. Phys. \textbf{17}, 515 (1976)
doi:10.1063/1.522935

\bibitem{Garfinkle:1993xk}
D.~Garfinkle, S.~B.~Giddings and A.~Strominger,
``Entropy in black hole pair production,''
Phys. Rev. D \textbf{49}, 958-965 (1994)
doi:10.1103/PhysRevD.49.958
[arXiv:gr-qc/9306023 [gr-qc]].

\bibitem{Dowker:1993bt}
F.~Dowker, J.~P.~Gauntlett, D.~A.~Kastor and J.~H.~Traschen,
``Pair creation of dilaton black holes,''
Phys. Rev. D \textbf{49}, 2909-2917 (1994)
doi:10.1103/PhysRevD.49.2909
[arXiv:hep-th/9309075 [hep-th]].

\bibitem{Dowker:1995gb}
F.~Dowker, J.~P.~Gauntlett, G.~W.~Gibbons and G.~T.~Horowitz,
``The Decay of magnetic fields in Kaluza-Klein theory,''
Phys. Rev. D \textbf{52}, 6929-6940 (1995)
doi:10.1103/PhysRevD.52.6929
[arXiv:hep-th/9507143 [hep-th]].

\bibitem{Gibbons:1987ps}
G.~W.~Gibbons and K.~i.~Maeda,
``Black Holes and Membranes in Higher Dimensional Theories with Dilaton Fields,''
Nucl. Phys. B \textbf{298}, 741-775 (1988)
doi:10.1016/0550-3213(88)90006-5

\bibitem{Costa:2000nw}
M.~S.~Costa and M.~Gutperle,
``The Kaluza-Klein Melvin solution in M theory,''
JHEP \textbf{03}, 027 (2001)
doi:10.1088/1126-6708/2001/03/027
[arXiv:hep-th/0012072 [hep-th]].

\bibitem{Gutperle:2001mb}
M.~Gutperle and A.~Strominger,
``Fluxbranes in string theory,''
JHEP \textbf{06}, 035 (2001)
doi:10.1088/1126-6708/2001/06/035
[arXiv:hep-th/0104136 [hep-th]].

\bibitem{Astorino:2012zm}
M.~Astorino,
``Charging axisymmetric space-times with cosmological constant,''
JHEP \textbf{06}, 086 (2012)
doi:10.1007/JHEP06(2012)086
[arXiv:1205.6998 [gr-qc]].

\bibitem{Lim:2018vbq}
Y.~K.~Lim,
``Electric or magnetic universe with a cosmological constant,''
Phys. Rev. D \textbf{98}, no.8, 084022 (2018)
doi:10.1103/PhysRevD.98.084022
[arXiv:1807.07199 [gr-qc]].

\bibitem{Horowitz:1998ha}
G.~T.~Horowitz and R.~C.~Myers,
``The AdS / CFT correspondence and a new positive energy conjecture for general relativity,''
Phys. Rev. D \textbf{59}, 026005 (1998)
doi:10.1103/PhysRevD.59.026005
[arXiv:hep-th/9808079 [hep-th]].


\bibitem{Kastor:2015wda}
D.~Kastor and J.~Traschen,
``Melvin Magnetic Fluxtube/Cosmology Correspondence,''
Class. Quant. Grav. \textbf{32}, no.23, 235027 (2015)
doi:10.1088/0264-9381/32/23/235027
[arXiv:1507.05534 [hep-th]].


\bibitem{Wald:1983ky}
R.~M.~Wald,
``Asymptotic behavior of homogeneous cosmological models in the presence of a positive cosmological constant,''
Phys. Rev. D \textbf{28}, 2118-2120 (1983)
doi:10.1103/PhysRevD.28.2118

\bibitem{Woolgar:2016axs}
E.~Woolgar,
``The rigid Horowitz-Myers conjecture,''
JHEP \textbf{03}, 104 (2017)
doi:10.1007/JHEP03(2017)104
[arXiv:1602.06197 [math.DG]].

\bibitem{Barzegar:2019vaj}
H.~Barzegar, P.~T.~Chru\'sciel, M.~H\"orzinger, M.~Maliborski and L.~Nguyen,
``Remarks on the energy of asymptotically Horowitz-Myers metrics,''
Phys. Rev. D \textbf{101}, no.2, 024007 (2020)
doi:10.1103/PhysRevD.101.024007
[arXiv:1907.04019 [gr-qc]].

\bibitem{El-Menoufi:2013pza}
B.~Mahmoud El-Menoufi, B.~Ett, D.~Kastor and J.~Traschen,
``Gravitational Tension and Thermodynamics of Planar AdS Spacetimes,''
Class. Quant. Grav. \textbf{30}, 155003 (2013)
doi:10.1088/0264-9381/30/15/155003
[arXiv:1302.6980 [hep-th]].


\bibitem{Kastor:2018cqc}
D.~Kastor, S.~Ray and J.~Traschen,
``Black Hole Enthalpy and Scalar Fields,''
Class. Quant. Grav. \textbf{36}, no.2, 024002 (2019)
doi:10.1088/1361-6382/aaf663
[arXiv:1807.09801 [gr-qc]].



\bibitem{Kastor:2006ti}
D.~Kastor and J.~Traschen,
``Stresses and Strains in the First Law for Kaluza-Klein Black Holes,''
JHEP \textbf{09}, 022 (2006)
doi:10.1088/1126-6708/2006/09/022
[arXiv:hep-th/0607051 [hep-th]].

\bibitem{El-Menoufi:2013tca}
B.~M.~El-Menoufi, B.~Ett, D.~Kastor and J.~Traschen,
``Sum Rule for the ADM Mass and Tensions in Planar AdS Spacetimes,''
Class. Quant. Grav. \textbf{30}, 205006 (2013)
doi:10.1088/0264-9381/30/20/205006
[arXiv:1305.1913 [hep-th]].

\bibitem{Kastor:2008wd}
D.~Kastor, S.~Ray and J.~Traschen,
``The Thermodynamics of Kaluza-Klein Black Hole/Bubble Chains,''
Class. Quant. Grav. \textbf{25}, 125004 (2008)
doi:10.1088/0264-9381/25/12/125004
[arXiv:0803.2019 [hep-th]].

\bibitem{Tian:1986zz}
Q.~Tian,
``Cosmic strings with cosmological constant,''
Phys. Rev. D \textbf{33}, 3549-3555 (1986)
doi:10.1103/PhysRevD.33.3549

\bibitem{Linet:1986sr}
B.~Linet,
``The static, cylindrically symmetric strings in general relativity with cosmological constant,''
J. Math. Phys. \textbf{27}, 1817-1818 (1986)
doi:10.1063/1.527050

\bibitem{AKST}
M.~Astorino, D.~Kastor, W.~Sabra and J.~Traschen, work in progress.

\bibitem{Sabra:2020gio} 
  W.~A.~Sabra,
  ``Kasner Branes with Arbitrary Signature,''
  arXiv:2005.03953 [hep-th].
  
\bibitem{Kastor:2016bnm}
D.~Kastor, S.~Ray and J.~Traschen,
``Genuine Cosmic Hair,''
Class. Quant. Grav. \textbf{34}, no.4, 045003 (2017)
doi:10.1088/1361-6382/aa5735
[arXiv:1608.04641 [hep-th]].


\bibitem{Gumrukcuoglu:2007bx}
A.~E.~Gumrukcuoglu, C.~R.~Contaldi and M.~Peloso,
``Inflationary perturbations in anisotropic backgrounds and their imprint on the CMB,''
JCAP \textbf{11}, 005 (2007)
doi:10.1088/1475-7516/2007/11/005
[arXiv:0707.4179 [astro-ph]].

\bibitem{Blanco-Pillado:2015dfa}
J.~J.~Blanco-Pillado and M.~Minamitsuji,
``The Initial State of a Primordial Anisotropic Stage of Inflation,''
JCAP \textbf{06}, 024 (2015)
doi:10.1088/1475-7516/2015/06/024
[arXiv:1501.07427 [hep-th]].



%
%
%





%
%


\end{thebibliography}
\end{document}